\documentclass[aps,prl,twocolumn,showpacs,superscriptaddress,groupedaddress]{revtex4}
\usepackage[caption=false]{subfig}
\usepackage{graphicx}  
\usepackage{dcolumn}   
\usepackage{bm}        
\usepackage{amssymb}   

\usepackage{titlesec}

\titleformat {\section}[block]
{\vspace{-15pt}\normalfont\itshape}{Appendix~\thesection:~}
{\vspace{-12pt}}
{}

\newcommand{\beq}{\begin{equation}}
\newcommand{\eeq}{\end{equation}}
\newcommand{\beqn}{\begin{eqnarray}}
\newcommand{\eeqn}{\end{eqnarray}}
\newcommand{\beqns}{\begin{eqnarray*}}
\newcommand{\eeqns}{\end{eqnarray*}}
\newcommand{\bei}{\begin{itemize}}
\newcommand{\eei}{\end{itemize}}

\newcommand{\rar}{\rightarrow}

\newcommand{\mumu}{\mu^+\mu^-}
\newcommand{\pipi}{\pi^+\pi^-}

\def\GeVM{~${\rm GeV}/c^2$}

\def\degree{$^\circ$}

\def\fitsa{\gamma_\mathrm{ISR}\gamma_\mathrm{SA}}
\def\fitla{\gamma_\mathrm{ISR}\gamma_\mathrm{LA}}
\def\fittsa{\gamma_\mathrm{ISR}\raisebox{-2pt}{2}\gamma_\mathrm{SA}}
\def\fitsala{\gamma_\mathrm{ISR}\gamma_\mathrm{SA}\gamma_\mathrm{LA}}
\def\fittla{\gamma_\mathrm{ISR}\raisebox{-2pt}{2}\gamma_\mathrm{LA}}
\def\thtrkla{\theta_{\min(\mathrm{trk},\gamma_\mathrm{LA})}}

\input babarsym

\begin{document}

\begin{flushleft}
\begin{minipage}{5cm}
BABAR-PUB-23/005\\
SLAC-PUB-17736 \\
\end{minipage}
\end{flushleft}

\title{Measurement of additional radiation in the initial-state-radiation processes $e^+e^-\rar\mu^+\mu^-\gamma$ and $e^+e^-\rar\pi^+\pi^-\gamma$ at \babar}

\author{J.~P.~Lees}
\author{V.~Poireau}
\author{V.~Tisserand}
\author{E.~Grauges}
\author{A.~Palano}
\author{G.~Eigen}
\author{D.~N.~Brown}
\author{Yu.~G.~Kolomensky}
\author{M.~Fritsch}
\author{H.~Koch}
% ...
\collaboration{The \babar\ Collaboration}
\noaffiliation

\begin{abstract}
A dedicated measurement of additional radiation in $e^+e^-\rar\mu^+\mu^-\gamma$ and $e^+e^-\rar\pi^+\pi^-\gamma$ initial-state-radiation events is presented using the full \babar\ data sample. For the first time results are presented at next-to- and next-to-next-to-leading order, with one and two additional photons, respectively, for radiation from the initial and final states.  
Comparison with predictions from \textsc{Phokhara} and \textsc{AfkQed} Monte Carlo generators is performed, revealing discrepancies in the one-photon rates and angular distributions for the former.  
This disagreement has a negligible effect on the \babar\ measurement of the $e^+e^-\rar\pi^+\pi^-(\gamma)$ cross section, but could affect other measurements significantly. This study sheds a new light on the longstanding discrepancy in this channel that affects the theoretical prediction of hadronic vacuum polarization contributions to the muon magnetic moment anomaly. 

\end{abstract}

\maketitle

\textit{Introduction.} 
The current Standard Model prediction~\cite{Aoyama:2020ynm} for the muon magnetic moment anomaly $a_\mu$ falls short of the direct
measurement by 5.0$\sigma$~\cite{Muong-2:2021ojo,Muong-2:2023cdq}, possibly indicating physics beyond the Standard Model. A leading source of uncertainty in $a_\mu$ arises from imprecise knowledge of the contribution of the hadronic vacuum polarization (HVP) evaluated from dispersion relations and measurements of
the $e^+e^- \to \mathrm{hadrons}$ cross section. The dominant channel in this cross section is $e^+e^- \to \pi^+\pi^-(\gamma)$,
which furnishes 73\% of the HVP contribution to $a_\mu$.
However, a longstanding discrepancy 
persists between the two most precise results using the initial-state radiation (ISR) technique, from \babar~\cite{BaBar:2009wpw,BaBar:2012bdw} and KLOE~\cite{KLOE08,KLOE10,KLOE12,KLOEcomb}. The result from KLOE lies 2.9$\sigma$ below that of \babar\ in the region of the $\rho$ resonance. Recent results from the CMD-3 experiment~\cite{CMD-3:2023alj} with the energy scan technique lie above both \babar\ and KLOE, exceeding them by 2.2$\sigma$ and 5.1$\sigma$, respectively. This situation prevents a meaningful update of the $a_\mu$ prediction to compare with the new Fermilab measurement~\cite{Muong-2:2023cdq}. Tension is also observed with the lattice QCD evaluation~\cite{Borsanyi:2020mff,rcqcd-window,window-Mainz22,window-ETMC22,window-RBC-UKQCD23,window-FHM23} of the HVP contribution. The need to clarify these tensions and improve the precision on $a_\mu$ calls for further studies of the different approaches.

In this Letter we focus on higher-order radiative processes, likely to affect 
differently the various ISR experiments, namely the next-to-leading order (NLO) processes $e^+e^-\to\mu^+\mu^-\gamma\gamma$ and $\pi^+\pi^-\gamma\gamma$, and the next-to-next-to-leading order (NNLO) processes $e^+e^-\to\mu^+\mu^-\gamma\gamma\gamma$ and $\pi^+\pi^-\gamma\gamma\gamma$. Each photon can be emitted from the initial-state $e^+$ or $e^-$, or a final-state muon or pion (FSR).
Building on our previous work~\cite{BaBar:2009wpw,BaBar:2012bdw}, where very loose kinematical constraints were applied in order to reduce dependence on simulation, we make improved measurements of the relative contributions of events with two photons and study their kinematics. We then perform the first study of events with three photons.

\textit{Dataset and simulated samples.} 
The analysis is based on $424.2\invfb$ ($43.9\invfb$) of data~\cite{lumi} collected 
at the SLAC PEP-II asymmetric $e^+e^-$ collider operated at (and 40\mev below) the $\FourS$ resonance. The \babar\ detector and performance are described in detail elsewhere~\cite{detector,performance}, 
and its response, including the variations of beam and detector conditions over time, is simulated with {\small GEANT4}~\cite{geant4}.

The two Monte Carlo (MC) generators used in this analysis are described in Appendix~A.
Signal processes $e^+e^-\to X\gamma(\gamma)$, with $X=\mu\mu$ or $\pi\pi$, 
are simulated with \textsc{Phokhara}  
with ten times the data statistics. 
Small samples of these processes are also simulated with the \textsc{AfkQed} generator. 
In both cases, one high-energy photon is generated within the detector angular range.
\textsc{Phokhara} generation is limited to NLO, with up to one additional ISR or FSR photon emitted with an angular distribution according to the full NLO matrix element. \textsc{AfkQed} provides NLO and NNLO simulated samples, within the approximation that the additional ISR photons are generated collinear to the beams.

Background ISR processes $e^+e^-\to K^+K^-\gamma(\gamma)$ and $e^+e^-\to X\gamma$ 
($X =\pi^+\pi^-\pi^0, \pi^+\pi^-2\pi^0$, ...) are simulated with 
\textsc{Phokhara} and 
\textsc{AfkQed}, respectively.
Background processes $e^+e^-\to\qqbar$ 
($q=u,d,s,c$) are generated with {\small JETSET}~\cite{jetset} 
and $e^+e^-\to\tau^+\tau^-$ with {\small KK2f}~\cite{kk2f}. 
As the models used in the generators for the background processes may be unreliable at low multiplicity or do not reproduce the measured hadronic spectra, data-driven corrections are applied to the MC estimations (Appendix~B).  
Backgrounds mostly affect the pion channel while they are very low in the muon channel, mainly from well simulated $\tau\tau$ events. 

\begin{figure}[ht]
\centering
\includegraphics[width=0.95\columnwidth]{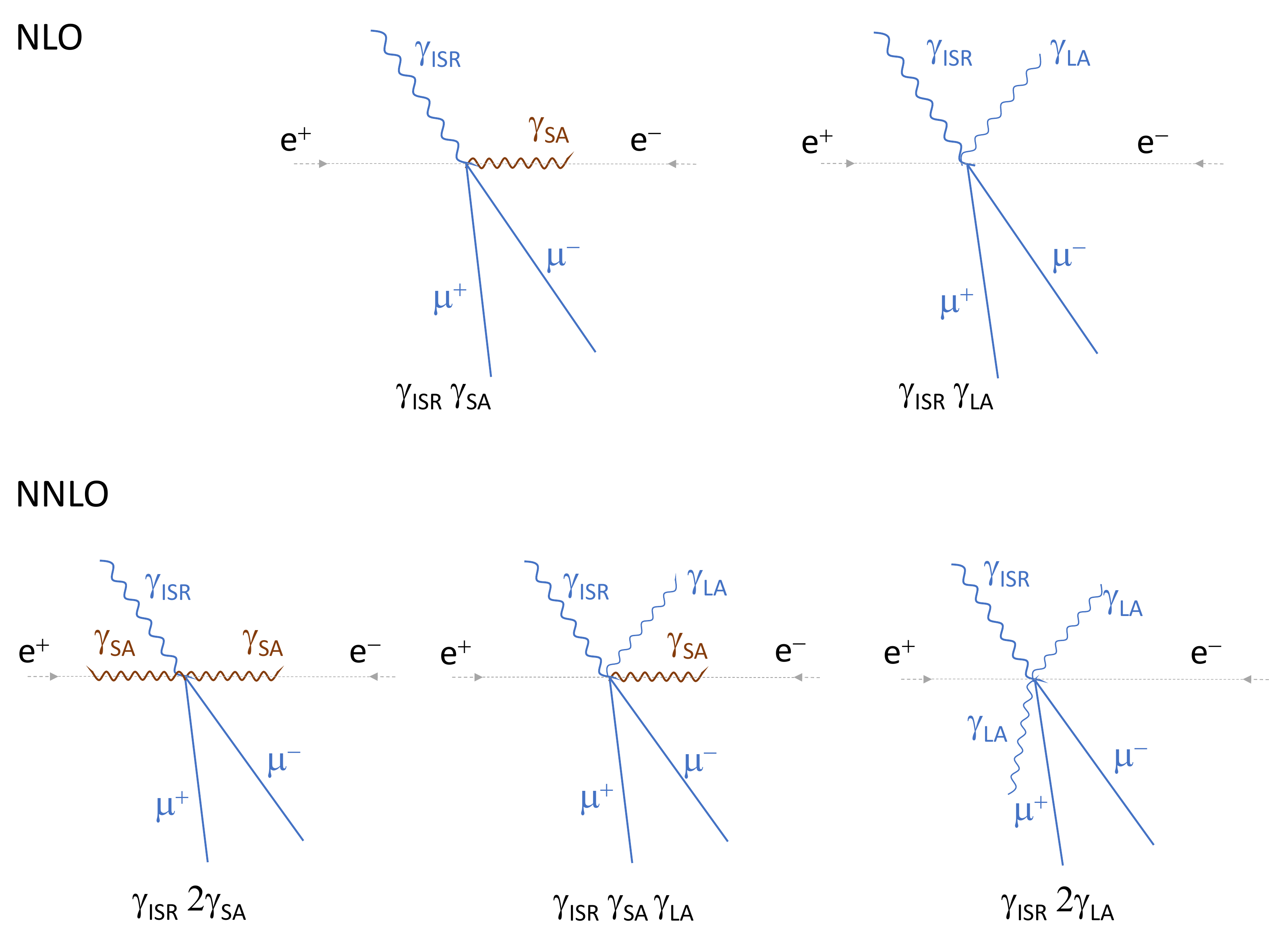}
\caption{ 
Sketch of measured topologies for the ISR $\mu\mu\gamma$ (or $\pi\pi\gamma$) process at NLO and NNLO levels. Tracks and photons in blue are measured in the detector, undetected photons in red are assumed to be aligned with a beam.}
\label{fig:sketch}
\end{figure}

\textit{Event selection.}
Radiative events 
are selected by requiring a detected photon in the laboratory polar-angle range $0.35<\theta<2.4$\rad,
and with a measured energy $E^\ast_\gamma>4\gev$ in the center-of-mass (c.m.) frame. 
Exactly two tracks of opposite charges are required to 
extrapolate to the collision region, each with: transverse momentum $> 0.1\gevc$, $\theta$ in the range 0.4$-$2.45\rad, at least 15 hits in the drift chamber.   
Events can contain any number of additional photons and tracks that do not satisfy the requirements.
Detected photons with $E_{\gamma}>50 \mev$ and polar angle in the range 0.35$-$2.4\rad are retained for the fits described below. 
The photon with the highest measured energy in the c.m.\ frame is denoted $\gamma_\mathrm{ISR}$.  
Events in which both tracks are tightly identified~\cite{performance} as muons (pions) are assigned to the dimuon (dipion) sample.
In the following, the rate of each radiative process is quoted as a fraction of the total yield in those dimuon and dipion samples, respectively.
Topologies considered in the analysis are sketched in Fig.~\ref{fig:sketch}.

\textit{NLO study.}
The dimuon and dipion samples are submitted to
two types of kinematic fits to 
$e^+e^-\to\pi^+\pi^-\gamma_\mathrm{ISR}\gamma$, with one photon in addition to $\gamma_\mathrm{ISR}$:
\begin{itemize}
\item[–] $\fitla$ fits:  a photon $\gamma_\mathrm{LA}$ is detected at large angle (LA) to the beam,
and its measured energy and angles are used in a fit.
\item[–] $\fitsa$ fits: an undetected small angle (SA)
photon $\gamma_\mathrm{SA}$ is approximated as being collinear with either beam. 
\end{itemize}

Both NLO fits use the measured $\gamma_\mathrm{ISR}$ 
energy and direction, and the  
parameters and covariance matrices of the two tracks. 
To be consistent with the cross section measurement, the pion mass is assumed for both tracks~\cite{BaBar:2009wpw,BaBar:2012bdw}.
All events undergo the $\fitsa$ fits while only events with at least one additional detected photon with $E_{\gamma}>50 \mev$ undergo the $\fitla$ fits, taking each photon in turn. 
The fit with the lowest $\chi^2$ is retained. 
An optimized contour in the ($\chi^2_{\fitsa},\chi^2_{\fitla}$) plane (so-called 2D-$\chi^2$ selection) 
is used to reduce backgrounds. 
This only selection applied to the NLO samples retains 99\% of the signal events, as determined in MC estimations and measured on the muon sample. 
Events are assigned to the NLO SA or LA samples according to the lowest $\chi^2$ and if the additional photon energy 
exceeds $0.2\gev$. As $\gamma_\mathrm{LA}$ photons are measured in the detector, the energy threshold is expressed in the laboratory system, while for $\gamma_\mathrm{SA}$ ISR photons, it is set in the c.m.\ frame. Events that satisfy the 2D-$\chi^2$, but not the energy, requirements are categorized as leading order (LO) events. 
We retain dimuons in the $m_{\pi\pi}$ mass range from threshold to 1.4\gevcc and dipions in the range between 0.6 and 0.9\gevcc around the $\rho$ resonance where backgrounds are manageable (5.5\% and 23\% for the SA and LA pion samples, respectively, dominated by muon misidentification). 
For the dipion sample, the background from fake photons due to pion interactions in the calorimeter is measured in data and MC simulations and the relative excess of $(21.5\pm 3.5)\%$ in data is corrected for.

\begin{figure}[ht]
\subfloat[]{\includegraphics[width=0.95\columnwidth]{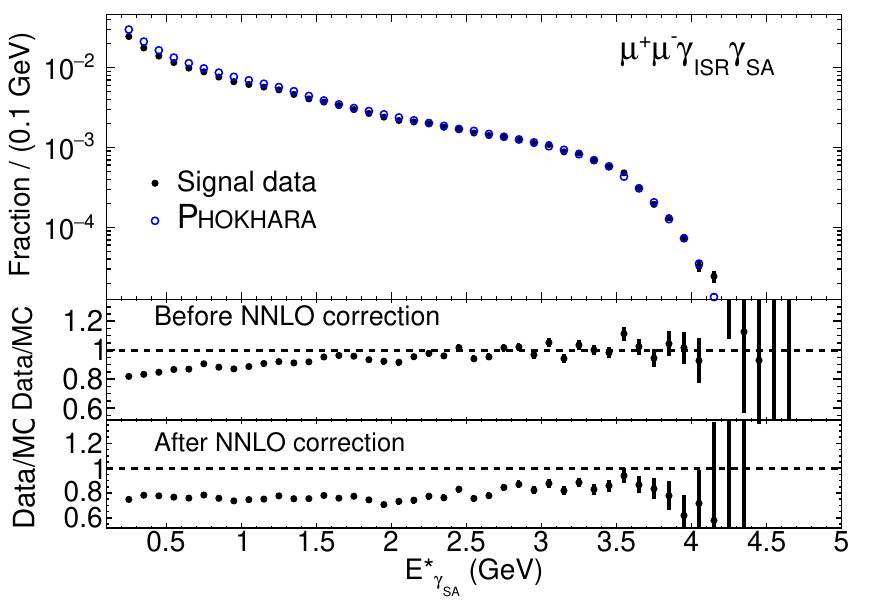}}\\
\subfloat[]{\includegraphics[width=0.95\columnwidth]{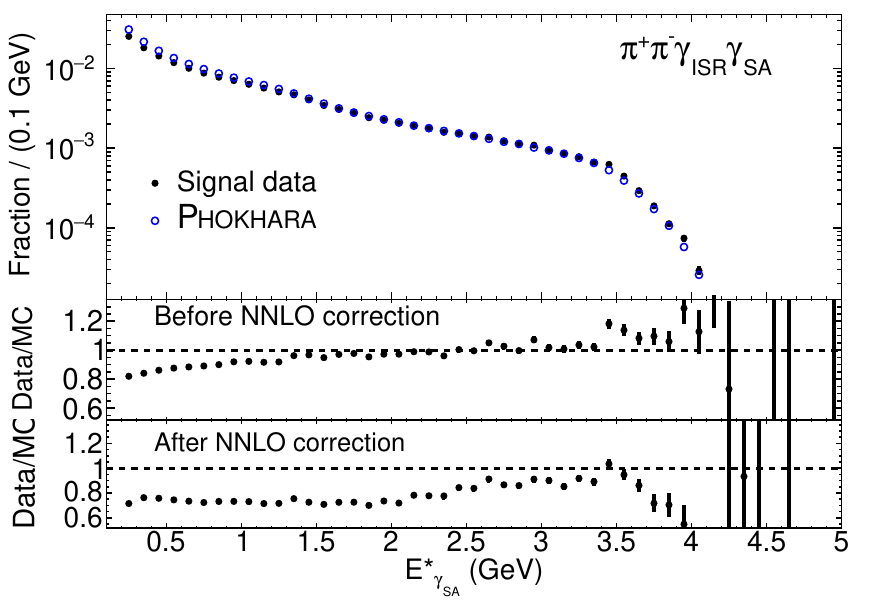}}
\vspace{-0.2cm}
\caption{
Distributions of the fitted energy of the SA photon in the c.m.\ frame for the muon (a) and pion (b) samples. The data (full dots) are compared with \textsc{Phokhara} (open/blue dots) in each upper plot, and their ratio is shown in the middle plot.  The lower plots show the ratio after the NNLO corrections discussed 
later in the text.
}
\label{fig:NLO} 
\end{figure}

In the SA samples, the energy distributions $E^\ast_{\gamma_\mathrm{SA}}$ of the additional photon are shown in Fig.~\ref{fig:NLO} for muons and pions.
As expected from their ISR nature, the results of the two processes are consistent. 
A marked discrepancy between data and the NLO \textsc{Phokhara} predictions is observed, both in shape and normalization, in the dominant low-energy region. 

\begin{figure}[ht]
\subfloat[]{\includegraphics[width=0.95\columnwidth]{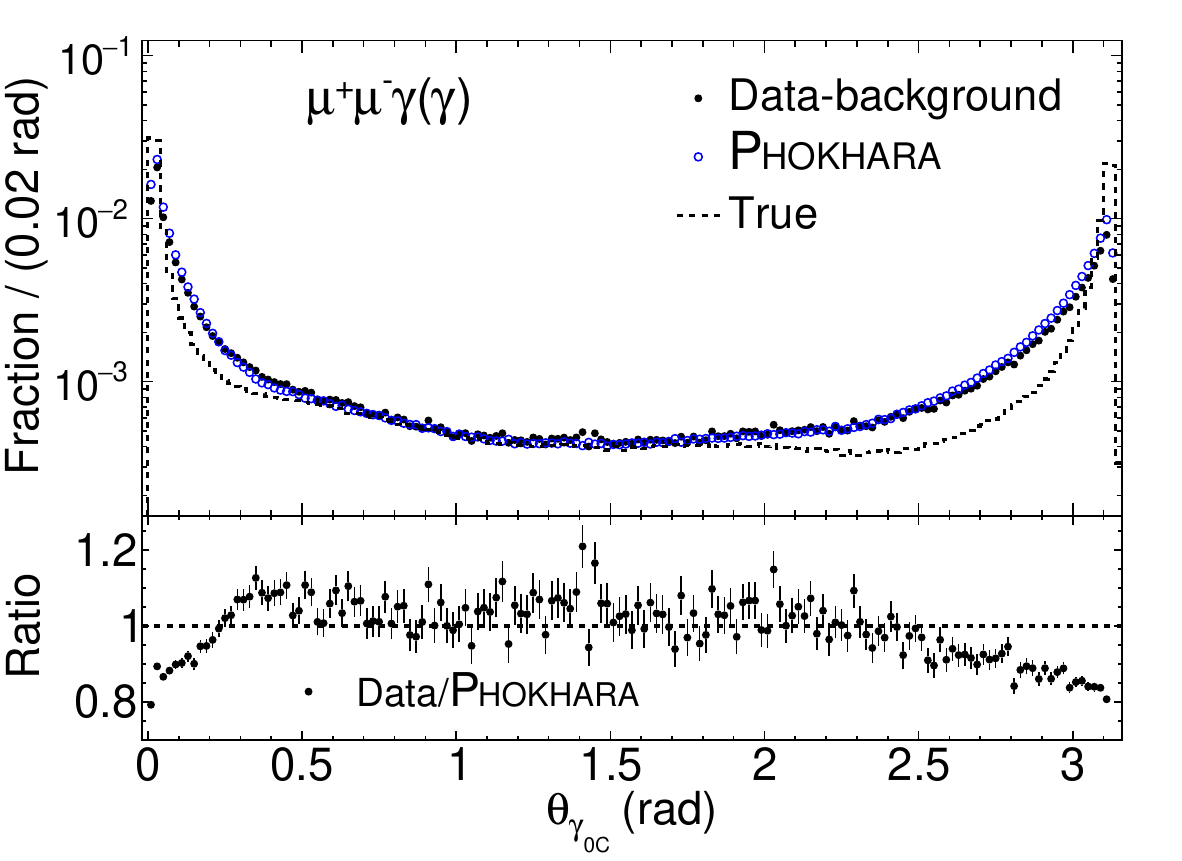}}\\
\subfloat[]{\includegraphics[width=0.95\columnwidth]{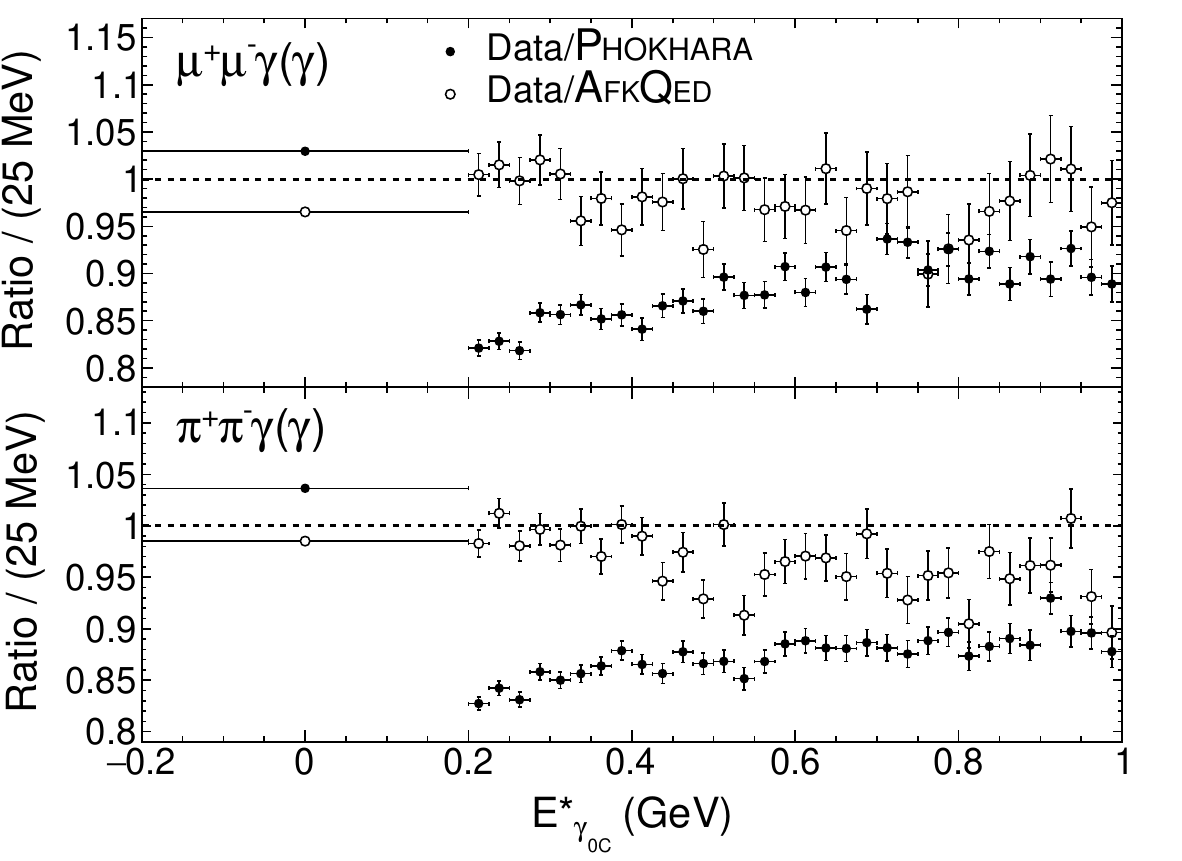}}
\vspace{-0.2cm}
\caption{\label{fig:0C} 
(a) Distributions of the calculated polar angle $\theta_{\gamma_\mathrm{0C}}$ 
for dimuon events with $E^\ast_{\gamma_\mathrm{0C}}>0.2\gev$ in the data (full dots) and \textsc{Phokhara} (open dots), along with the calculated $\theta_{\gamma_\mathrm{0C}}$ using MC truth information (dashed histogram).  (b)  Ratios of data to \textsc{Phokhara} (full dots) and data to \textsc{AfkQed} (open dots) distributions of $E^\ast_{\gamma_\mathrm{0C}}$ in the energy range below and above 0.2\,GeV corresponding to LO and NLO, respectively, in muon (upper) and pion (lower) samples.}
\end{figure}

This discrepancy is further investigated.
The collinearity approximation of the $\gamma_\mathrm{SA}$ photon with the beams 
in the $\fitsa$ fit is found to induce energy- and polar-angle-dependent biases for both the $\chi^2$ value and the $\gamma_\mathrm{SA}$ fitted energy, 
as determined in MC studies. However a comparison of the outputs of the $\fitsa$ and $\fitla$ fits for events with a detected photon verifies with data that the $E^\ast_{\gamma_\mathrm{SA}}$ bias is correctly simulated,
hence is not responsible for the observed data to \textsc{Phokhara} discrepancy.
A further test, independent from the NLO fits, 
is provided by a zero-constraint (0C) calculation of 
the energy and angles of the additional photon. 
With the constraint of 4-momentum conservation in $e^+e^-\to \pi^+\pi^-{\gamma_\mathrm{ISR}}{\gamma_\mathrm{0C}}$,
the calculation uses the two track momenta and the $\gamma_\mathrm{ISR}$ direction. 
To avoid spurious $\gamma_\mathrm{0C}$ photons arising from shower fluctuations,
the ${\gamma_\mathrm{ISR}}$ energy measurement is ignored 
and $\gamma_\mathrm{0C}$ candidates within a cone of 0.5\rad around the $\gamma_\mathrm{ISR}$ direction are excluded. 
The distributions of the
polar angle $\theta_{\gamma_\mathrm{0C}}$ for events with a calculated photon energy $E^\ast_{\gamma_\mathrm{0C}}$ above 0.2 \gev, are shown in Fig.~\ref{fig:0C}(a) for the data and \textsc{Phokhara} dimuon samples. Comparison reveals that \textsc{Phokhara} severely overestimates the NLO rate at small angles to the beams, with improving agreement at larger angles. 
As detailed studies demonstrate (Appendix~C), this feature is not due to $\theta_{\gamma_\mathrm{0C}}$ resolution.
The data/\textsc{Phokhara} ratio of the photon energy spectra $E^\ast_{\gamma_\mathrm{0C}}$, shown in Fig.~\ref{fig:0C}(b), reveals a mismatch between the LO and NLO rates and presents 
a positive slope, as in the $\fitsa$ fit. 
Since no angular assumption enters the 0C calculation, the observed energy dependence of the data/\textsc{Phokhara} ratio 
is thus not an artifact caused by the collinearity approximation of the $\gamma_{\mathrm{SA}}$ photon.
In contrast to \textsc{Phokhara}, neither mismatch between the LO and NLO categories nor a data/MC energy slope are observed when conducting the same exercise with \textsc{AfkQed}.

 In the LA samples, no energy dependence is observed in the data/MC ratio of the $E_{\gamma_\mathrm{LA}}$ distributions from the $\fitla$ fit, neither with \textsc{Phokhara} nor \textsc{AfkQed},
 and the data to \textsc{Phokhara} discrepancy in normalization is much reduced.
 The relative contributions of ISR and FSR processes are measured using the minimum of the two angles in the laboratory frame between a track and the LA photon $\thtrkla$~\cite{BaBar:2012bdw}. 
 We observe clear peaks in the data and MC distributions of this angle below 20\degree, due to FSR, over a broad distribution from LA ISR (Appendix~D).  
 
\textit{NNLO study.}
We then perform 
an NNLO analysis using three kinematic fits, $\fittsa$, $\fitsala$ and $\fittla$, where two photons are allowed in addition to $\gamma_\mathrm{ISR}$.
The SA photons $\gamma_\mathrm{SA}$ are approximated to be collinear with one of the beams and the LA photons $\gamma_\mathrm{LA}$ are measured in the detector.
All events undergo the $\fittsa$ fit, with one SA photon along each beam. Only events with at least one or two additional detected photons with $E_{\gamma}>50 \mev$ 
enter the $\fitsala$ and $\fittla$ fits, respectively. 
No 2D-$\chi^2$ selection is applied, and events are assigned to the $\fittsa$, $\fitsala$ or $\fittla$ sample if the $\chi^2$ of that fit is lower than all others, including the NLO fits.

\begin{figure*}[t]
\subfloat[]{\includegraphics[width=0.95\columnwidth]{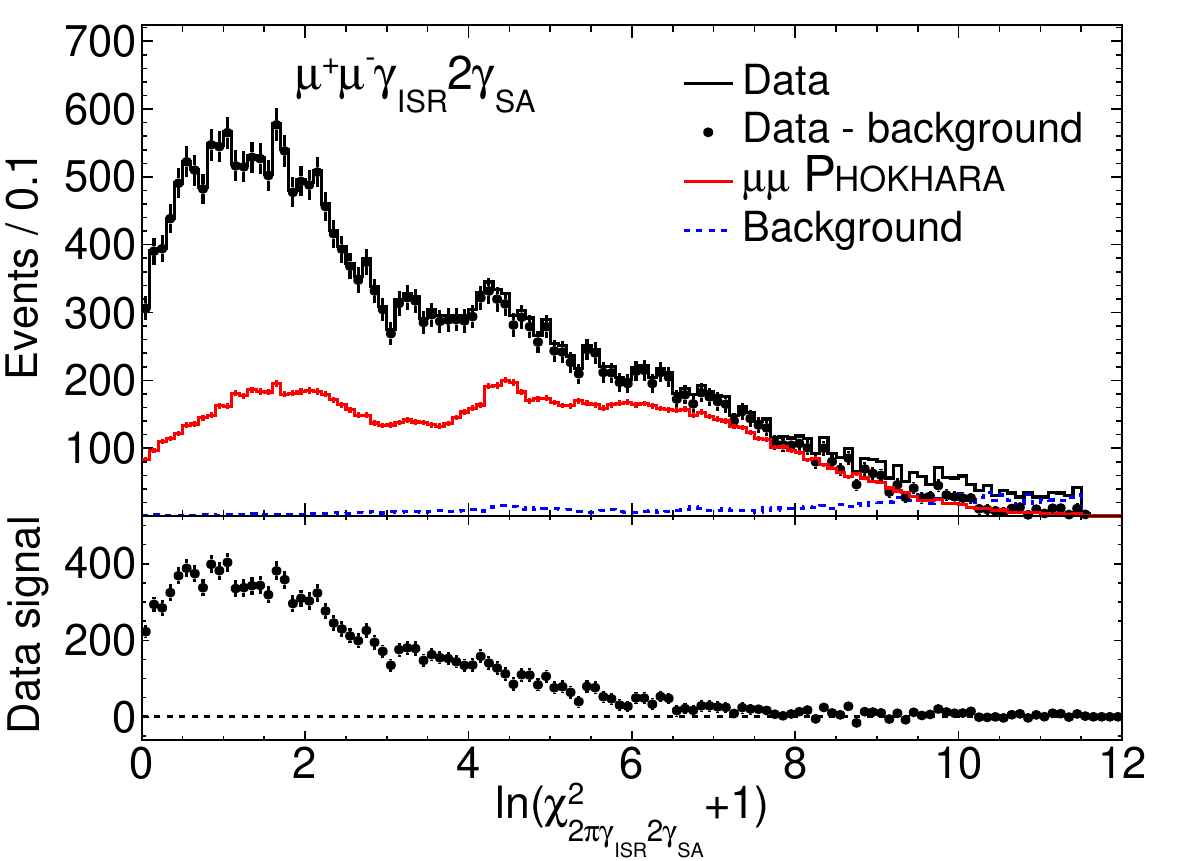}}
\subfloat[]{\includegraphics[width=0.95\columnwidth]{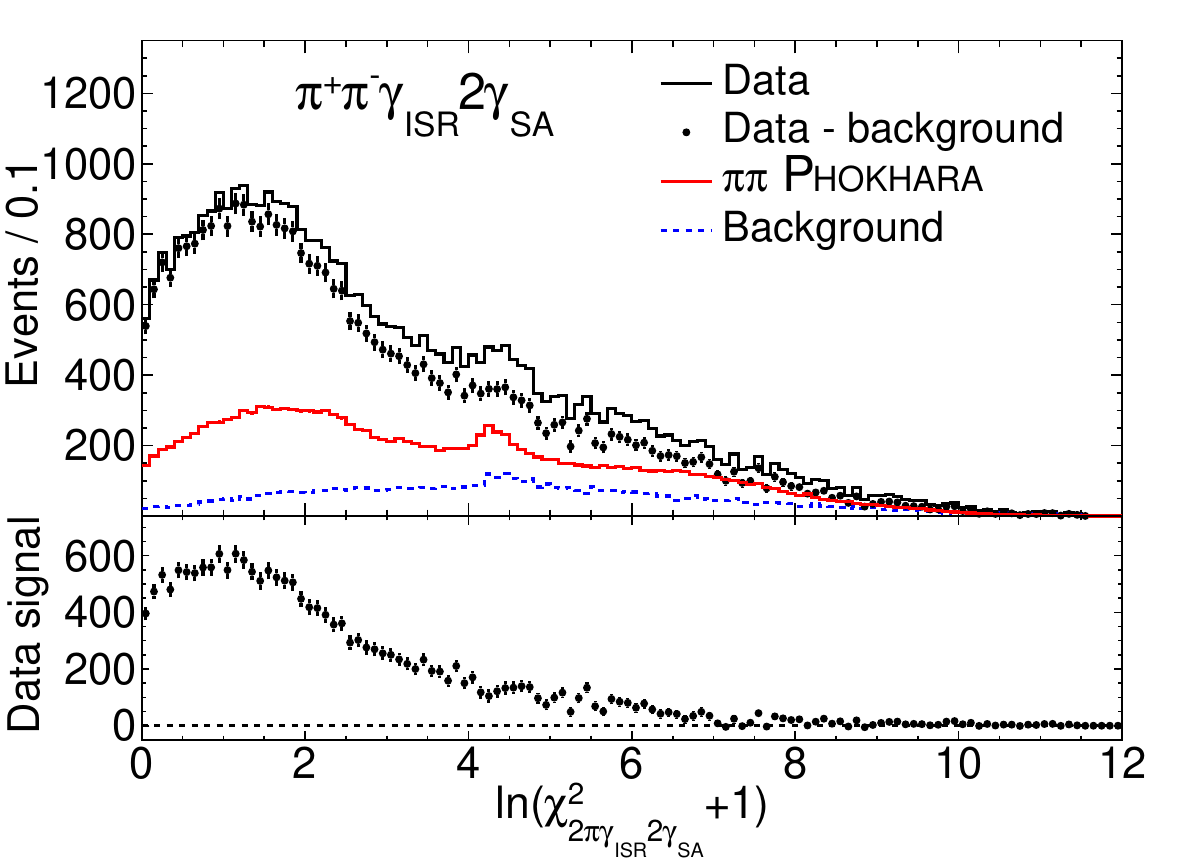}}\\
\subfloat[]{\includegraphics[width=0.95\columnwidth]{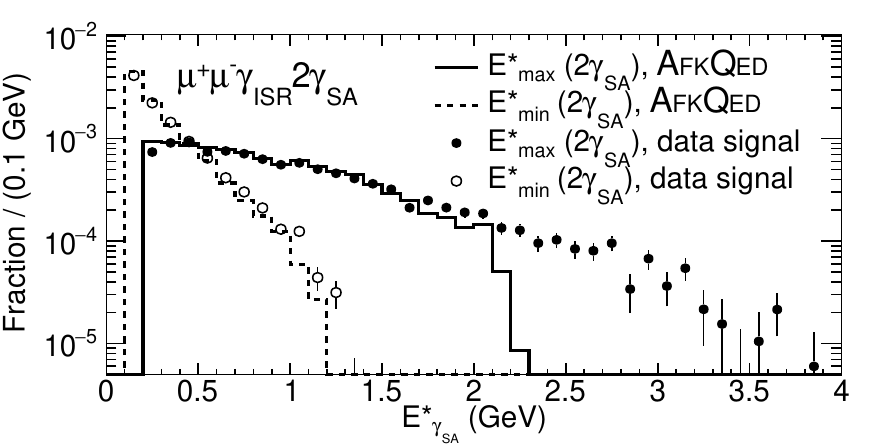}}
\subfloat[]{\includegraphics[width=0.95\columnwidth]{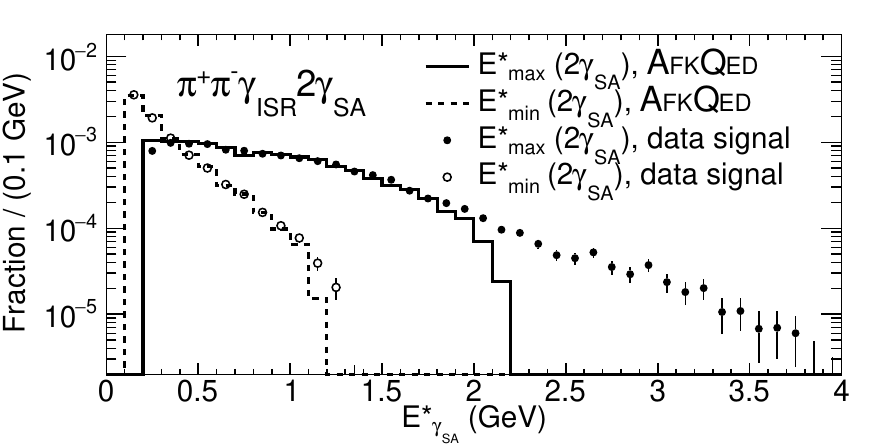}}
\vspace {-0.2cm}
\caption{\label{fig:NNLO} Distributions of the $\chi^2$ (a), (b) of the $\gamma_\mathrm{ISR}2\gamma_\mathrm{SA}$ kinematic fit for the muon (left) and 
pion (right) samples. The corresponding distributions (c), (d) of the c.m.\ energies of the more (full dots) and less energetic (open dots) SA photons 
after subtraction of backgrounds and NLO feedthrough, compared with \textsc{AfkQed} predictions (histograms), 
normalized to the data for $E^\ast_{\gamma_\mathrm{SA}}<2.3\gev$.}
\end{figure*}

In all NNLO categories, the background  
from NLO events with spurious additional photons
is estimated with the \textsc{Phokhara} sample, after correcting for data and MC rate differences observed above at LO and NLO levels. 
The non-NLO background is very small in the dimuon samples, but the dipion samples are affected by larger background from multihadronic ISR processes, especially those containing a $\pi^0$.
This background is reduced by BDT (boosted decision tree) selections designed for each NNLO category (Appendix~E).
To further reduce background contributions, the fitted additional photons 
are required to pass energy thresholds specific to each NNLO category.
After these selections, the NLO 
contribution to the background is dominant in all NNLO muon categories and pion $\fittsa$.
In the pion $\fitsala$ and $\fittla$ samples, the 
multihadronic ISR contributions dominate but do not exceed 45\% of the data.
The NNLO data signal yields are obtained by subtracting both the NLO contribution and the estimated non-NLO background. 

Significant NNLO signals are found in data in all 
$\fittsa$, $\fitsala$ and $\fittla$ categories.
Results for the most abundant $\fittsa$ category are summarized in Fig.~\ref{fig:NNLO}, 
requiring $E^\ast_{\gamma_\mathrm{SA}}>0.2\gev$ for the most energetic SA photon and $E^\ast_{\gamma_\mathrm{SA}}>0.1\gev$ for the other.
The $\chi^2$ distributions and energy spectra of the SA photons
are shown for muons (a), (c) and pions (b), (d). 
In both samples, the NLO background (solid/red histogram) dominates over the very low non-NLO 
background (dashed/blue histogram).
The energy spectra of the SA photons for $\fittsa$ signals in data 
are further compared to \textsc{AfkQed} predictions, where the \textsc{AfkQed} NNLO signals are restricted to the events with two SA photons that satisfy the energy thresholds. 

\textit{Final results combining NLO and NNLO studies.}
Final results are obtained considering all NLO and NNLO radiative processes
using a feedthrough probability matrix between true and reconstructed categories established with the \textsc{Phokhara} and \textsc{AfkQed} samples. 
Feedthrough contributions are estimated through an iterative procedure to correct for rate differences of each category in data and MC samples.  
Feedthrough yields estimated with \textsc{AfkQed} are corrected to account for its missing LA-ISR component.

The NLO $\fitsa$ rates and $E^\ast_{\gamma_\mathrm{SA}}$ energy distributions 
are further corrected for a hidden NNLO feedthrough of the same size as the open $\fittsa$ rate,
due to the additional collinear ISR radiation: while 
double SA radiation from opposite beams is identified by the $\fittsa$ fit,
double SA radiation from the same beam cannot be distinguished from a single-photon radiation 
where the equivalent single photon  
merges the two $\gamma_\mathrm{SA}$ energies.
The same-beam NNLO component in the measured $\fitsa$ spectrum is thus inferred from the sum of $\gamma_\mathrm{SA}$ energies in data $\fittsa$ events, and subtracted.
As shown in the bottom panels in Fig.~\ref{fig:NLO}, the NLO SA photon energy spectrum in data after the NNLO correction agrees better in shape with the \textsc{Phokhara} prediction, but it has a deficit in rate by more than 20\%.

\begin{table}[htbp]
\centering
\caption{Event fractions in data for the $\mu\mu$ and $\pi\pi$ processes in all fit categories. The numbers in parentheses represent uncertainties, where the first is statistical and the second systematic.
The results, except for NNLO 2LA (which is not simulated by any 
generator currently available) are corrected using efficiencies that vary category by category between 
$99\%$ and $72\%$, except for NLO FSR $\pi\pi$ ($40\%$) and NNLO FSR $\pi\pi$ ($22\%$
due to BDT selection.)}
\label{tab:fraction-final-data}
\begin{tabular}{l|l|l} \hline
 Category     & \hspace{8mm}$\mu\mu$ & \hspace{10mm}$\pi\pi$\\
              & $m_{\pi\pi}<1.4\gevcc$ & $0.6<m_{\pi\pi}<0.9\gevcc$\\\hline
LO            & \hspace{3mm}0.7716(4)(14) & \hspace{3mm}0.7839(5)(12) \\

NLO SA-ISR    & \hspace{3mm}0.1469(3)(36) & \hspace{3mm}0.1401(2)(16) \\

NLO LA-ISR    & \hspace{3mm}0.0340(2)(9)  & \hspace{3mm}0.0338(2)(9) \\

NLO ISR       & \hspace{3mm}0.1809(4)(35)    & \hspace{3mm}0.1739(3)(20) \\

NLO FSR       & \hspace{3mm}0.0137(2)(7)    &  \hspace{3mm}0.0100(1)(16) \\

NNLO ISR~\footnote{\scriptsize NNLO ISR = 2SA-ISR or SA-ISR\,+\,LA-ISR} & \hspace{3mm}0.0309(2)(38)   &  \hspace{3mm}0.0310(2)(39) \\

NNLO FSR~\footnote{\scriptsize NNLO FSR = SA-ISR\,+\,LA-FSR}      & \hspace{3mm}0.00275(6)(9)   &  \hspace{3mm}0.00194(12)(50)\\

NNLO 2LA~\footnote{\scriptsize NNLO 2LA = 2LA-ISR, LA-ISR\,+\,LA-FSR or 2LA-FSR}      & \hspace{3mm}0.00103(3)(1) & \hspace{3mm}0.00066(4)(4) \\\hline

\end{tabular}
\end{table}

The final fractions of all fit categories with respect to the total event sample are given in Table~\ref{tab:fraction-final-data}. 
The systematic uncertainties include those from efficiency corrections, background subtraction and feedthrough corrections with the latter two being the dominant contributions.
Template fits to the $\thtrkla$ distributions in the $\fitla$, $\fitsala$ and $\fittla$ samples are used to estimate the fractions of FSR and large-angle ISR at the NLO and NNLO levels (Appendix~D).

The results are summarized below:
\begin{itemize}
\item 
    NNLO contributions are clearly observed
    with a total fraction of 
    $(3.47\pm0.38)$\% for muons and $(3.36\pm0.39)$\% for pions. 
    This allows the correction of NLO rates for cross feeds from NNLO categories.  The corrected shape of the energy distribution for single SA photons provides further evidence, and good internal consistency.
    
\item 
    The NLO SA-ISR fractions in the \textsc{Phokhara} generator are higher than in the data, with data/MC ratios of $0.763\pm0.019$ for dimuons and $0.750\pm0.008$ for dipions, while the respective LA-ISR ratios, $0.96\pm0.03$ and $0.98\pm0.03$, are consistent with unity.    
    This indicates a problem in the angular distribution of the NLO photon generated by \textsc{Phokhara}, with a large excess at small angles to the beams.

\item 
    The \textsc{AfkQed} generator provides a reasonable description of the rates and energy distributions
    of NLO and NNLO data.
    The sum of SA-ISR and LA-ISR rates in data up to the cutoff at 2.3\gev photon energy applied at generation (Appendix~A) leads to slightly high data/MC ratios of  
    $1.061\pm0.015$
    for muons and $1.043\pm0.010$ for pions. 

\item 
The ratio between data and the \textsc{Phokhara} prediction for NLO FSR is found to be $0.86\pm0.05$ for muons and $0.76\pm0.12$ for pions. 
The corresponding data/\textsc{AfkQed} ratios are $1.09\pm0.06$ 
for muons and $1.08\pm0.10$ for pions.
In both cases, the ratio pion/muon is consistent with unity and supports the pion pointlike behavior for additional FSR. 
This result, obtained with a fake-photon subtraction that both includes a
data/MC correction and takes into account NNLO feedthrough, supersedes the previous result of an excess of $(21 \pm 5)$\% reported in
the \babar\ publication~\cite{BaBar:2009wpw,BaBar:2012bdw}.

\end{itemize}

\textit{Consequences for the $\pi^+\pi^-(\gamma)$ cross section ISR measurements.}
The consequences of these findings for the $\mu^+\mu^-(\gamma)$ and $\pi^+\pi^-(\gamma)$ cross section measurements depend strongly on the experimental approach.

The \babar\ measurements~\cite{BaBar:2009wpw,BaBar:2012bdw} 
are performed with a very loose selection 
that incorporates all NLO and higher order radiative processes. 
Therefore, no dependence on a particular event generator is introduced 
when handling additional radiation (Appendix~F). 
Only the acceptance, determined using \textsc{Phokhara}, 
is affected by the generator shortcomings.
However the acceptance correction of $(0.3 \pm 0.1)\times 10^{-3}$ is negligible compared to the estimated systematic uncertainty of 0.5\% in the $\pi^+\pi^-(\gamma)$ cross section measurement~\cite{BaBar:2009wpw,BaBar:2012bdw}.

In contrast, other experiments using the ISR approach~\cite{KLOE08,KLOE10,KLOE12,BESIII,CLEOc} 
do not measure additional radiation. They select events mostly in the LO topology, which includes virtual and soft emission, and rely on the \textsc{Phokhara} generator to account for the missing hard NLO contribution. 
While the total NLO correction in the cross section is strongly reduced by the near cancellation of the  
hard NLO terms by the infrared-finite sum of soft and virtual terms~\cite{KLN}, 
each component is large and systematic uncertainties related to the undetected hard NLO contribution far exceed the theoretical uncertainty in the total NLO correction.
The results of the present analysis question the validity of the procedure relying on \textsc{Phokhara} on two grounds: first, the hard NLO yield in \textsc{Phokhara} is significantly larger than our measurement,
and second, NNLO contributions, absent in \textsc{Phokhara}, are found to be at a level larger than the systematic uncertainties quoted in Refs.~\cite{KLOE08, KLOE10, KLOE12, BESIII, CLEOc}.
A quantitative evaluation of these effects requires detailed studies of the specific experimental conditions,
which is beyond the scope of this Letter.
However the unique measurements reported here, and their implication for ISR experiments that rely on the  \textsc{Phokhara} generator to account for unmeasured radiative events, 
offer new key insights that may allow progress in the longstanding quandary that affects the theoretical prediction of HVP contributions to $a_\mu$, with respect to both the final value and its currently degraded precision. 

\vspace{12pt}
\noindent \textit{Acknowledgments.} We are grateful for the extraordinary contributions of our PEP-II colleagues in achieving the excellent luminosity and machine conditions that have made this work possible. The success of this project also relies critically on the expertise and dedication of the computing organizations that support \babar, including GridKa, UVic HEP-RC, CC-IN2P3, and CERN. The collaborating institutions wish to thank SLAC for its support and the kind hospitality extended to them. We also wish to acknowledge the important contributions of J.~Dorfan and our deceased colleagues E.~Gabathuler, W.~Innes, D.W.G.S.~Leith, A.~Onuchin, G.~Piredda, and R. F.~Schwitters.
This work is supported by the ``ADI 2020'' project funded by the IDEX Paris-Saclay, ANR-11-IDEX-0003-02 and NSFC (Grant No.\ 11975153).


\section*{Appendix A: Signal Monte Carlo generators.} 
\vspace{-5mm}
In this analysis, initial-state-radiation events $e^+e^- \to X\gamma$ are produced by Monte Carlo generators with a high-energy photon emitted at large angle to the beams.
In \textsc{Phokhara9.1}~\cite{phokhara-v9.1} complete NLO corrections to $e^+e^- \to \mu^+\mu^-\gamma$ and $e^+e^- \to \pi^+\pi^-\gamma$ are implemented. All NLO diagrams for virtual and real photon production are included, whether radiation occurs in initial or final state. For muons, matrix elements are given by QED for ISR, FSR, and interference. For pions, ISR radiation is given by QED while FSR is described by a pointlike model of the pion. No NNLO diagram is included. 
Up to one additional ISR or FSR photon is generated with an angular distribution according to the full NLO matrix element.

The \textsc{AfkQed} generator is
based on the formalism of Refs.~\cite{eva,eva2}. 
Additional ISR photons are generated collinear to the $e^+$ or $e^-$ beams
with the structure function method~\cite{struct-fct}. 
Additional FSR radiation is implemented at NLO with \textsc{Photos}~\cite{photos}.
As the structure function method relies on a resummation of leading logarithms, it takes into account real and virtual higher-order contributions, although in an approximate way.
\textsc{AfkQed} hence provides simulated samples with explicit higher order topologies, such as one additional ISR photon and one additional FSR photon or multiple ISR emission occurring on both beams.
In the samples used in this analysis, a minimum mass $m_{X\gamma}>8\gevcc$ imposed at generation
limits the additional ISR energy to 2.3\gev. 

\section{Appendix B: Background normalization.}\label{app-bkg}
\vspace{-5mm}
The main backgrounds in the pion channel are $e^+e^-\to\qqbar$, dominant at high mass, and $e^+e^-\to \pi^+\pi^-\pi^0\gamma$ dominant at low mass. 
Their estimates are normalized using the data in the 2D-$\chi^2$ selected region, where they amount to 0.7\% and 2.4\% of the expected signal, respectively.  
From the comparison of data and simulated yields of $\pi^0$ that mimic an ISR photon, the $uds$ background is scaled by a factor $0.422\pm0.025$. 
For the $3\pi$ channel dominated by the production of $\omega$ and $\phi$ resonances, normalization is obtained from reconstructing $\pi^0$ from non-ISR photons and comparing the resonance peaks in the $\pi^+\pi^-\pi^0$ spectrum in data and MC samples. A scale factor of $1.013\pm 0.021$ is applied to the $3\pi\gamma$ MC.
For the other ISR background channels, weights are applied to the MC events depending on the hadronic mass to correct for differences between MC mass distributions and spectra measured at \babar.   

\section{Appendix C: Angular resolution of the 0C calculation.}\label{app-0Cresol}
\begin{figure}[h]
\includegraphics[width=0.95\columnwidth]{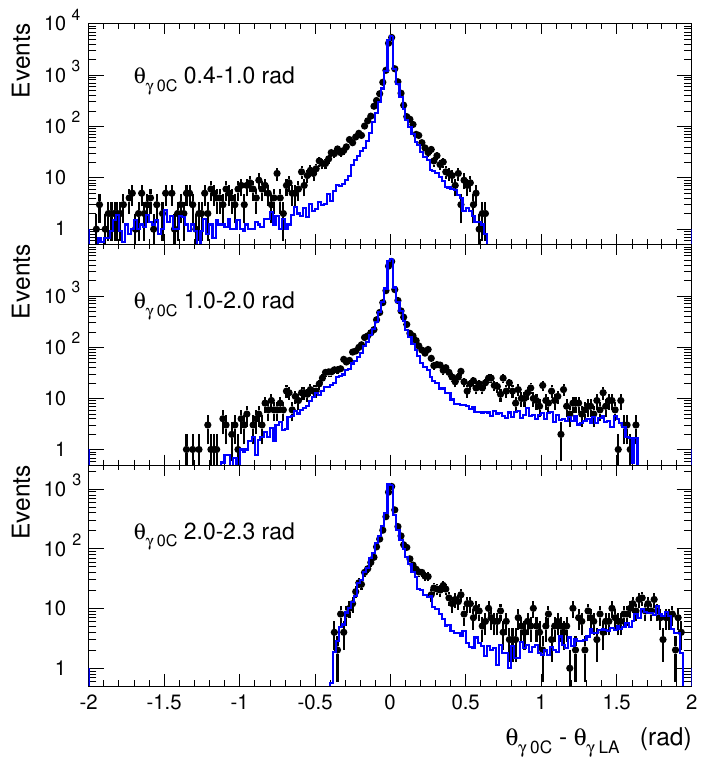}
\caption{\small Distribution of angular differences between the calculated angles $\theta_{\gamma_\mathrm{0C}}$ and the measured one $\theta_{\gamma_\mathrm{LA}}$ from the $\fitla$ fit, in three $\theta_{\gamma_\mathrm{0C}}$ intervals in which the additional photon is within the angular detector acceptance: data points and MC histograms (blue).}
\label{fig:0Cresol}
\end{figure}

\vspace{-5mm}
Resolution of $\theta_{\gamma_\mathrm{0C}}$ determination is estimated on MC events by comparing the value calculated from truth information with the value calculated from the reconstructed quantities. As seen in Fig.~\ref{fig:0C}(a), resolution induces a reconstructed/true ratio much larger than unity at low angles, hence opposite to the data/\textsc{Phokhara} deficit.   
The angular resolution is estimated in muon data and MC by comparing the calculated polar angle $\theta_{\gamma_\mathrm{0C}}$ with the fitted one from the $\fitla$ fit, as shown in Fig.~\ref{fig:0Cresol} for $E^\ast_{\gamma_\mathrm{0C}}>0.2\gev$.
There is good agreement between data and simulation for the core of the resolution function, with rms values of 30\mrad with little dependence on 
$\theta_{\gamma_\mathrm{0C}}$, but tails above 0.5\rad  
are more important in data. The resulting larger transfer of photons from the dominant sharp radiation peaks along the incident beams towards large angles 
is estimated to enhance the data/MC ratio by $\sim$10\% in the central $\theta_{\gamma_\mathrm{0C}}$ region, as observed in Fig.~\ref{fig:0C}(a).

\section{Appendix D: FSR and LA-ISR separation.}\label{app-FSR}
\begin{figure}[h]
\includegraphics[width=0.95\columnwidth]{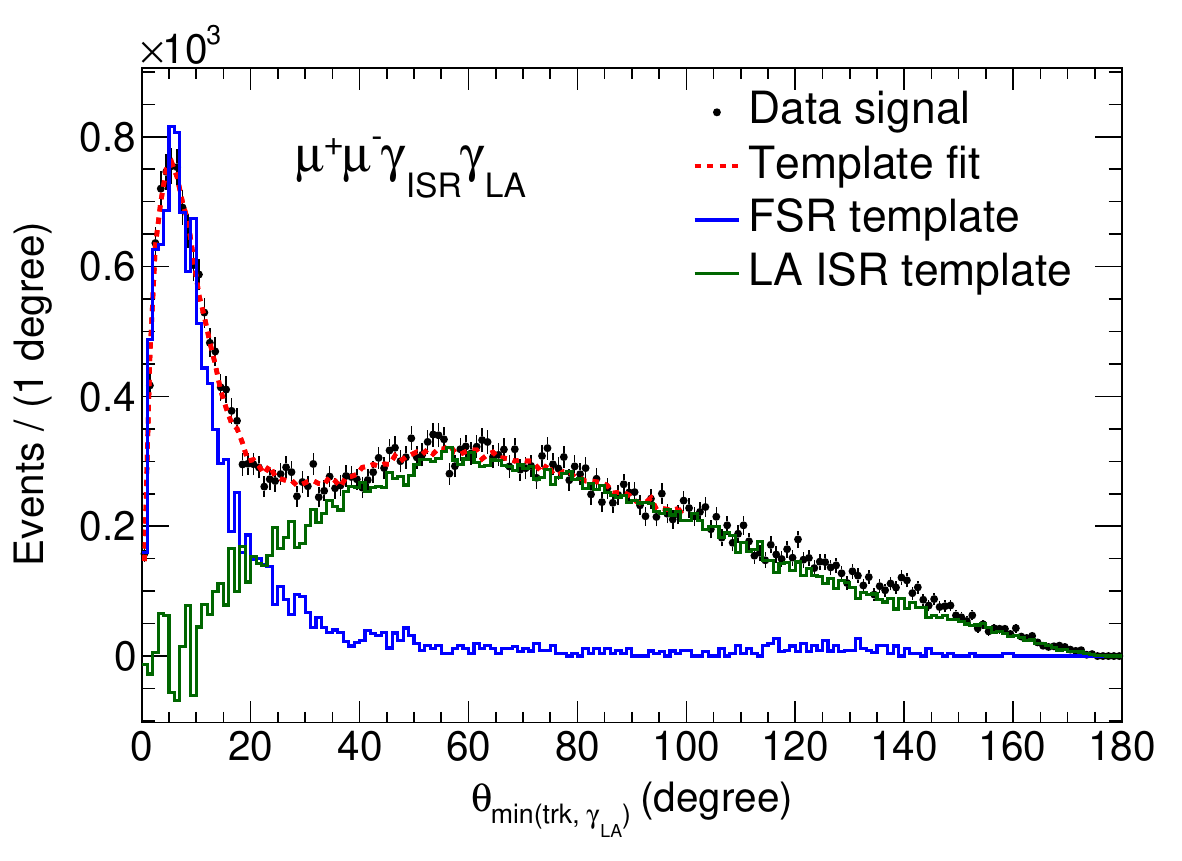}
\caption{\small Template fit separating FSR events from LA ISR events in the $\mu\mu\fitla$ data sample.}
\label{fig:fsr-laisr}
\end{figure}

\vspace{-5mm}
The distribution of the minimum angle $\thtrkla$ between the LA photon and one of the two tracks is shown in Fig.~\ref{fig:fsr-laisr} for the $\mu\mu\fitla$ data sample. The two components are fitted using FSR and LA-ISR templates. As LA-ISR is absent in \textsc{AfkQed}, the LA photons are due uniquely to FSR and their $\thtrkla$ distribution provides the FSR template. In \textsc{Phokhara} both components are present and the LA-ISR template is obtained by subtracting the FSR template normalized to the yield below $10^\circ$ from the total distribution. The template fit of the data $\thtrkla$ distribution allows us to measure the relative rates of the two components. Template fits are similarly applied in the $\pi\pi\fitla$ sample and to $\thtrkla$ distributions in $\fitsala$ and $\fittla$ samples.

\section{Appendix E: BDT selections.}\label{app-BDT}
\begin{figure}[htbp]
\centering
\includegraphics[width=\columnwidth]{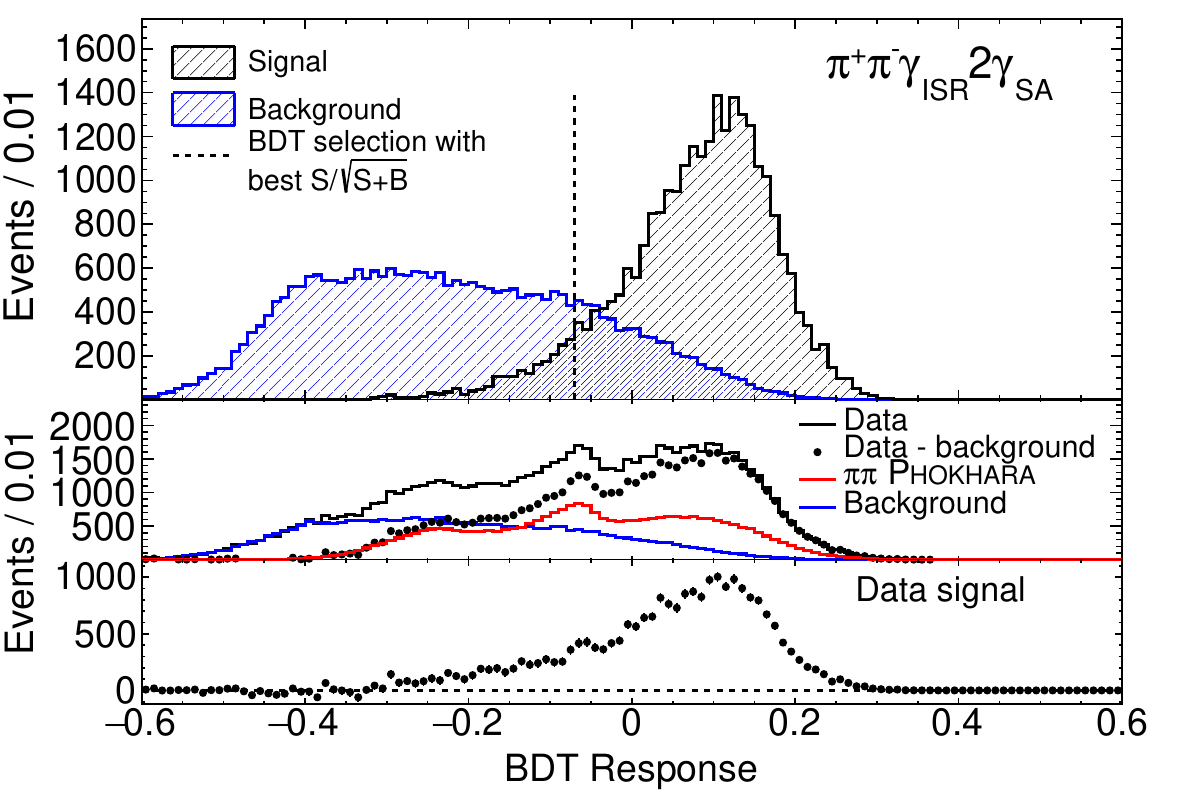}
  \caption{\label{bdt-1g2bm-pipi} \small BDT response distributions for the training samples (top panel);
   BDT response distributions showing data and different background components (middle and bottom panels). }
\end{figure}

\vspace{-5mm}
BDT techniques are used at several steps of this analysis: i) to select clean $uds$ and ISR $2\pi\pi^0$ samples to properly normalize these backgrounds; ii) to optimize the 2D-$\chi^2$ selection; iii)  to reduce backgrounds in NNLO pion samples. 
As an example, the BDT designed to select the $\fittsa$ pion signal uses discriminant variables as the $\gamma_\mathrm{ISR}$ energy and angles with respect to other photons and tracks, multiplicity and total energy of additional photons, and the pion angle with respect to the ISR photon direction in the dipion c.m.\ system. The signal training sample is from the NNLO \textsc{AfkQed} events satisfying the $\chi^2$ and SA photon energy requirements of the $\fittsa$ selection, while background is from simulation of non-$\pi\pi$ processes. Figure~\ref{bdt-1g2bm-pipi} shows the corresponding BDT response distributions for signal and backgrounds (top), and the comparison between data and backgrounds (bottom). The good agreement of the data and non-$\pi\pi$ background in the far-negative BDT response region indicates that the background estimation is satisfactory. The response of the NLO background taken from \textsc{Phokhara} is also well simulated. The selection applied maximizes the signal-over-background ratio and results in a BDT efficiency of $0.926(2)$ (where the uncertainty is statistical).

\section{Appendix F: The BaBar data-driven approach.}\label{app-bbr}
\vspace{-5mm}
Previous and ongoing measurements of the $\pi\pi$ cross section in \babar\ rely on NLO kinematic fits that explicitly include one photon in addition to $\gamma_\mathrm{ISR}$. Although some higher order events may be rejected by the NLO fits, this loss is corrected for by the 2D-$\chi^2$ efficiency measured on the data~\cite{BaBar:2009wpw,BaBar:2012bdw}.
It is only when calculating the event acceptance prior to the kinematic fits that generator dependence occurs. 

\begin{figure}[htbp] 
\centering
  \includegraphics[width=0.95\columnwidth]{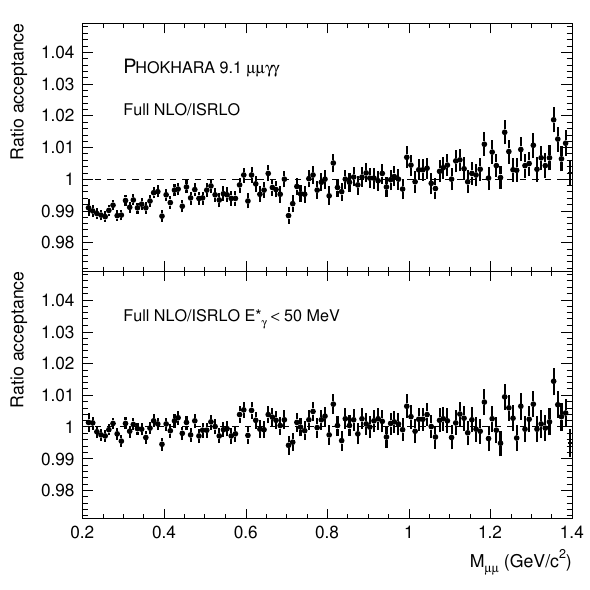}
  \caption{\label{fig:ratio-NLOsoft-LO} \small Ratio of the \babar\ acceptance for full NLO to LO acceptance for $\mu\mu\gamma(\gamma)$ \textsc{Phokhara} events. 
  Top: full NLO. Bottom: virtual+soft NLO ($E^*_{\gamma}<$50\mev). }
\end{figure}

To quantify the effect of the overestimated hard NLO contribution in \textsc{Phokhara}, the \babar\ acceptance is computed as a function of mass at LO and NLO levels using different options of the generator. Figure~\ref{fig:ratio-NLOsoft-LO} shows the ratio of NLO/LO acceptances, for full NLO events (top) or events with hard NLO ($E^\ast_{\gamma}> 50\mev$) excluded (bottom). Final states at LO (no generated additional photon) have the same acceptance as virtual+soft NLO events (no hard additional photon) at better than 1 per mil [Fig.~\ref{fig:ratio-NLOsoft-LO}(bottom)]. 
In contrast, hard NLO radiation does affect the acceptance, 
with a small variation in mass within 1\%  [Fig.~\ref{fig:ratio-NLOsoft-LO}(top)]. 
This acceptance correction, which needs to be computed from \textsc{Phokhara}, is however strongly correlated between the $\pi\pi\gamma(\gamma)$ and $\mu\mu\gamma(\gamma)$ processes, and therefore the effect largely vanishes when taking their ratio for the $\pi\pi$ cross section measurement, for an overall acceptance correction of $(0.9981\pm0.0004)$, constant with mass from threshold up to 1.4\GeVM.
The systematic bias induced 
by the hard NLO excess observed in \textsc{Phokhara} and the missing hard NNLO component is estimated from this result to be $(0.3 \pm 0.1)\times 10^{-3}$.

\end{document}